\documentclass[
twocolumn,secnumarabic
,amssymb, amsmath,nobibnotes, aps, prl]{revtex4}
\usepackage{graphicx}
\usepackage{tabularx}
\usepackage{bm}%
\expandafter\ifx\csname package@font\endcsname\relax\else
 \expandafter\expandafter
 \expandafter\usepackage
 \expandafter\expandafter
 \expandafter{\csname package@font\endcsname}%
\fi

\begin{document}

\title{Wilsonian Renormalization Group for a Multitrace Matrix Model}

\author{ Badis Ydri$^{a}$, Rachid Ahmim$^{b}$}
\affiliation{$^{a}$Department of Physics, Badji-Mokhtar Annaba University,\\
 Annaba, Algeria\\
$^{b}$Physics Department, Hamma-Lakhdar El Oued University,\\
El Oued, Algeria.
}
%\date{}

\begin{abstract}

 The Wilsonian renormalization group approach to matrix models is outlined and applied to multitrace matrix models with emphasis on the computation of the fixed points which could describe the phase structure of noncommutative scalar phi-four theory.

\end{abstract}

\maketitle
\tableofcontents

\section{Introduction}

    The renormalization group equation is a mathematical framework (very suited for answering foundational questions) as well as an efficient machinery for explicit non-perturbative calculations (of the same power as the Monte Carlo method) which gives quantum field theory its substance and its predictive power.

    The exact meaning of the renormalization group process may however differ across disciplines and authors but Wilson's approach remains the most influential not to mention the most profound and at the same time most intuitive of all (see for example \cite{Wilson:1973jj}).

    In here we will appropriate this understanding to elucidate the meaning of the renormalization group as well as its precise use which are the most useful to our purposes.

    We have a physical system characterized by some field variable $\Phi$ together with a partition function $Z$ which depends on a set of coupling constants $\{g\}$ and on a cut-off $\Lambda$, i.e. $Z=Z_{\Lambda}(g)$. In our case  the field variable $\Phi$ is an $N\times N$ hermitian matrix with action $NS(\Phi)=NTr_NV(\Phi)$. The typical example is quantum gravity in two dimensions as given in terms of quantum surfaces \cite{DiFrancesco:1993cyw} but more importantly, for us in this note, a quantum field theory over a noncommutative space.

    The Wilsonian renormalization group approach consists thus in the following very reasonable assumptions:
    \begin{itemize}
    \item The physical system is characterized by a very large but finite number of degrees of freedom ${\cal N}=N^2$. In this case $N$ acts as the cut-off $\Lambda$ or equivalently as an inverse lattice spacing $1/a$. This is a physical cut-off here arising from the underlying Planck structure of Euclidean spacetime.
    \item We reduce the density of degrees of freedom of the system by integrating out some high energy modes. In the case of matrix models this is done by decomposing the $N\times N$ matrix $\Phi$ into an $(N-1)\times(N-1)$ matrix $\phi$,  two complex $(N-1)-$vectors $v$ and $v^{\dagger}$ and a scalar $\alpha$ \cite{Brezin:1992yc} (see also \cite{Zinn-Justin:2014wva}). We write then 

\begin{eqnarray}
\Phi=\left( \begin{array}{cc}
\phi & v  \\
v^{\dagger} & \alpha \end{array} \right).
\end{eqnarray}
The action decomposes then as $NS(\Phi)=NS(\phi)+N\delta V(\phi)$ where $\delta V(\phi)$ is the quantum potential.
\item The fundamental assumption behind the renormalization group approach is the fact that the total free energy of the system ${\cal F}(N,g)=N^2F_N=-\ln Z_N$, which is the most basic physical property of the system, will not change under the change of the scale $N\longrightarrow N^{\prime}=N-1$.

  In other words, the free energy is assumed to remain constant under the process of integrating out "the high energy modes" $v$, $v^{\dagger}$ and $\alpha$ (together with an appropriate rescaling of  the "low energy mode" $\phi$). This is achieved by assuming that the coupling constants $g$ depend themselves on the cut-off $N$ and thus any change in the scale $N\longrightarrow N^{\prime}$ will also cause a response in the form of a change in the coupling constants as $g\longrightarrow g^{\prime}$ in such a way that the free energy of the system ${\cal F}$ remains constant.
   
  \item We have then
  \begin{eqnarray}
    {\cal F}(N,g)={\cal F}(N^{\prime},g^{\prime}).
  \end{eqnarray}
  It is in the sense that the cut-off $\Lambda=N$ is thought of to be unphysical.  In some sense the system does not change under any reduction of the density of degrees of freedom (which is intimately tied to scale invariance, renormalizability and conformal field theory).

  The above equation leads directly to a  highly non-linear renormalization group equation of the form 
\begin{eqnarray}
  (N\frac{\partial }{\partial N}+2)F(N,g)=G(g,a)~,~a=\frac{\partial }{\partial g}F(N,g).
\end{eqnarray}
The variable $a$ is the moment associated with the coupling constant $g$. This equation should be compared with the usual Callan-Symanzik renormalization group equation
\begin{eqnarray}
  (N\frac{\partial }{\partial N}+\gamma(g))F(N,g)=r(g)+\beta(g)\frac{\partial }{\partial g}F(N,g).
\end{eqnarray}
In our case the scaling dimenison is $\gamma=2$ whereas the linear behavior in $F$ of the right-hand side is replaced with a highly non-linear behavior encoded in the function $G(g,a)$.
\item Explicitly, the function $G$ is given in terms of the quantum potential by
  \begin{eqnarray}
    G&=&\frac{1}{N}\langle V(\phi)\rangle+\langle \delta V(\phi)\rangle.
  \end{eqnarray}
  The quantum potential will be dominated by a saddle point and by means of the loop equation the function $G$ can be expressed in terms  of the resolvent.
\item The linear term in the expansion of the function $G$ in powers of the moment $a$ depends on the beta functions 
\begin{eqnarray}
    g-g^{\prime}\equiv \frac{\partial g}{\partial N}=\frac{1}{N}\beta(g,N).
  \end{eqnarray}
By integrating this equation we get the renormalization group flow of the coupling constants $g$ as functions $g=g(N)$ of $N$. These functions represent surfaces in the space $g-N$ along which the free energy ${\cal F}$ is constant. Hence, the renormalization group equation (starting from some initial condition) moves us along a surface of constant ${\cal F}$ in the space $g-N$ as we vary $N$.

\item The zero of the beta functions $\beta(g,N)$ is the fixed point $\{g_*\}$ of the renormalization group flow which is  defined by the equation
\begin{eqnarray}
    \beta(g_*,N)=0.
  \end{eqnarray}
Thus, for $N_1\neq N_2$ the renormalization group flow $g=g(N)$ determines two sets of values $\{g_1\}$ and $\{g_2\}$ of the coupling constants $\{g\}$ corresponding to the beta functions $\beta_1(g,N_1)$ and $\beta_2(g,N_2)$ which intersect only at the fixed point $\{g_*\}$.

\item Hence, the fixed point  $\{g_*\}$ is independent of $N$ and it is the point where a continuum limit can be constructed.  The free energy is non-analytic around the fixed point with a non-trivial suscpetibility exponent $\gamma$ which is determined by the usual equation
  \begin{eqnarray}
    \beta^{\prime}(g_*,N)=0.
  \end{eqnarray}
  \item Expansion of the free energy (in the planar limit) around the fixed point allows us to determine the fixed point $g_*$, the susceptibility exponent $\gamma_0$, the first and second moments $a_1$ and $a_2$. In particular, the most singular term leads after substitution in the expanded renormalization group equation  to the identity
\begin{eqnarray}
  0=\beta= \frac{\partial G}{\partial a}|_{g_*}.
\end{eqnarray}
\end{itemize}
In this note we will carry out this programme  for the case of a doubletrace cubic-quartic matrix model which captures the main features of the phase structure of noncommutative phi-four in dimension two including the uniform ordered phase \cite{Ydri:2017riq}. The multitrace matrix model approach to noncommutative scalar field theory was developed originally in  \cite{Saemann:2010bw,OConnor:2007ibg}.

This note is organized as follows. In section two we solve the renormalization group equations for the cubic and the cubic-quartic matrix models  \cite{Higuchi:1994rv}. In section three we present our first attempt at extending the formalism to multitrace matrix models of noncommutative quantum field theory by considering the example of the doubletrace matrix model proposed in \cite{Ydri:2017riq}. Section four contains a brief conclusion and outlook.

\section{Matrix renormalization group equation}
 We consider $N\times N$ hermitian matrices $\Phi$ with a potential energy given by (with $g_0=g_1=0$ and $g_2=1$)
   \begin{eqnarray}
    V(\Phi)=g_0+g_1\Phi+\frac{g_2}{2}\Phi^2+\frac{g_3}{3}\Phi^3+\frac{g_4}{4}\Phi^4+....
   \end{eqnarray}
   The partition function is defined by 
  \begin{eqnarray}
    Z_N=\int {\cal D}\Phi~\exp(-N Tr_N V(\Phi)).
  \end{eqnarray}
  The free energy $F$ of the $N\times N$ matrix $\Phi$ is given by the usual formula
  \begin{eqnarray}
    F_N=-\frac{1}{N^2}\ln Z_N.
  \end{eqnarray}
 The basic physical content of the renormalization group equation is the statement that the total free energy ${\cal F}=N^2F_N=-\ln Z_N$ of the physical system (here a matrix model representing quantum gravity in two dimensions or a noncommutative field theory) must be independent of $N$ (which acts thus as a cutoff). The renormalization group equation is of the general form  \cite{Higuchi:1994rv}
\begin{eqnarray}
  &&(N\frac{\partial }{\partial N}+2)F(N,g_i)=G(g_i,a_i)\nonumber\\
  &&a_i=\frac{\partial }{\partial g_i}F(N,g_i)=\frac{1}{jN}\langle Tr\Phi^j\rangle.
\end{eqnarray}
The function $G$ is linear in the beta functions  ${\beta}_n(g)\equiv \beta_{n1}(g)$ given by 
\begin{eqnarray}
{\beta}_n(g)=-N\frac{\partial g_n^{}}{\partial N}.
\end{eqnarray}
But, the linear dependence of the function $G$ on these beta functions  is only the first term in its expansion in powers of $a_n$, viz
\begin{eqnarray}
  G(g,a) &=&\beta_0(g)+\sum_{n=3}^{\infty}\sum_{k=1}^{\infty}\beta_{nk}(g)a_n^k.
\end{eqnarray}
For the cubic-quartic potential we compute the following function    \cite{Higuchi:1994rv,Higuchi:1993nq,Higuchi:1993tg,Fukuma:1990jw}
 \begin{eqnarray}
       G
       &=&(-\frac{1}{2}g_3a_3-g_4a_4+\frac{1}{2})\nonumber\\
       &+&V(\bar{\lambda})-2\ln\bar{\lambda}-2\int_{-\infty}^{\bar{\lambda}}dz(W(z)-\frac{1}{z}).\nonumber\\
  \end{eqnarray}
  The saddle point $\bar{\lambda}$ is the solution of the equation 
    \begin{eqnarray}
     V^{\prime 2}(\bar{\lambda})=4Q(\bar{\lambda}).
    \end{eqnarray}
    While $W(\bar{\lambda})$ is the expectation value of the resolvent which is given by the solution of the loop equation. Explicitly, we have
\begin{eqnarray}
W(\bar{\lambda})=\frac{1}{2}\big(V^{\prime}(\bar{\lambda})-\sqrt{V^{\prime 2}(\bar{\lambda})-4Q(\bar{\lambda})}\big).
\end{eqnarray}
The function $Q(\bar{\lambda})$ can be computed by means of the $n=0$ and $n=-1$ Schwinger-Dyson identities to be given by
\begin{eqnarray}
Q(\bar{\lambda})&=&1+g_3\big(a_1 +\bar{\lambda}\big)+g_4\big(2a_2 +\bar{\lambda}^2+\bar{\lambda} a_1\big).
\end{eqnarray}
By using also Schwinger-Dyson identities we can always re-express the linear and quadratic moments $a_1$ and $a_2$ in terms of higher moments (this is related to the fact that why we can always set $g_1=0$ and $g_2=1$). For example,  we have $a_2=-3g_3a_3/2-2g_4a_4+1/2$.

The fixed point $(g_{3*},g_{4*})$ is a simultaneous zero of the beta functions $\beta_3(g)$ and $\beta_4(g)$ or equivalently
      \begin{eqnarray}
\frac{\partial G}{\partial a_3}=0~,~\frac{\partial G}{\partial a_4}=0.
     \end{eqnarray}
      We find four solutions:
      \begin{itemize}
      \item The Gaussian fixed point $(g_{3*},g_{4*})=(0,0)$.
      \item The pure quantum gravity fixed point $(g_{3*},g_{4*})=(0,-1/12)$ corresponding to a $(2,3)-$minimal conformal matter coupled to two-dimensional quantum gravity (Liouville theory).
         \item Another quantum gravity fixed point $(1/432^{0.25},0)$ corresponding to another  theory of $(2,3)-$minimal conformal matter coupled to two-dimensional quantum gravity. This points admits also the interpretation of the Ising fixed point in noncommutative field theory.
        \item A fixed point $(g_{3*},g_{4*})=(0.3066,0.0253)$ corresponding to a $(2,5)-$minimal conformal matter coupled to two-dimensional quantum gravity.
        \end{itemize}
    
\section{Multitrace matrix models}
               As an example of multitrace matrix models of noncommutative field theory we consider the doubletrace matrix model (see \cite{Ydri:2015vba} and \cite{Tekel:2015uza,Tekel:2015zga,Subjakova:2020haa})
                \begin{eqnarray}
V=BTrM^2+CTrM^4+DTrMTrM^3.\label{BCD}
                \end{eqnarray}
                As it turns out, random multitrace matrix models such as (\ref{BCD})  allow for the emergence of quantum  geometry \cite{Ydri:2017riq,Ydri:2015zsa,Ydri:2016daf} in a mechanism very different from the usual mechanism of emergent noncommutative geometry obtained from Yang-Mills matrix models \cite{Delgadillo-Blando:2007mqd,Ydri:2016kua}.
                
                In the large $N$ limit (saddle point) the scaling of this potential is given by
                \begin{eqnarray}
                  \frac{V}{N^2}=\frac{\tilde{B}}{N}Tr\tilde{M}^2+\frac{\tilde{C}}{N}Tr\tilde{M}^4+\frac{\tilde{D}}{N^2}Tr\tilde{M}Tr\tilde{M}^3.
                \end{eqnarray}
                The scaled field is $\tilde{M}=N^{1/4}M$ whereas the scaled parameters are $\tilde{B}=B/N^{3/2}$, $\tilde{C}=C/N^2$ and $\tilde{D}=D/N$. The doubletrace term is of the same order as the quartic term and stability requires that $\tilde{C}>-\tilde{D}$. The partition function is given by
               \begin{eqnarray}
                 &&Z=\int {\cal D}M~\exp(-N Tr_{N} V(M))\nonumber\\
                 &&V(M)=\frac{g_2}{2}M^2+\frac{g_4}{4}M^4+\frac{g}{3N}(Tr_NM)M^3.
               \end{eqnarray}
               We have the results
               \begin{eqnarray}
                 g_2=\pm 1~,~g_4=\frac{\tilde{C}}{\tilde{B}^2}~,~g=\frac{3}{4}\frac{\tilde{D}}{{\tilde{B}^2}}.
               \end{eqnarray}
               The above doubletrace potential captures the essential features of the phase structure of noncommutative scalar phi-four theory in any dimensions which consists of three stable phases: i) disordered (symmetric, one-cut, disk) phase $\langle M\rangle =0$, ii) uniform ordered (Ising, broken, asymmetric one-cut) phase $\langle M\rangle\sim {\bf 1}$ and iii) non-uniform ordered (matrix, stripe, two-cut, annulus) phase $\langle M\rangle \sim \Gamma$ with $\Gamma^2={\bf 1}$.  See \cite{Ydri:2015vba} and references therein for example \cite{Shimamune:1981qf}.

               The 3rd order transition from the disordered phase to the non-uniform phase is the celebrated matrix transition \cite{Brezin:1977sv} which is expected to be described by the Gaussian fixed point of the quartic or cubic matrix model.
               
               The transition from disordered to uniform, which appears for small values of $C$ and negative values of $B$, is the celebrated 2nd order Ising phase transition and it is the one that is expected to be captured by the renormalization group equation since $a_1\neq 0$ in the uniform phase and $a_1|_{a_3=0}= -g_3$ in the pure cubic matrix model. 
               
\begin{figure}[htbp]
\begin{center}
  \includegraphics[width=10cm,angle=-0]{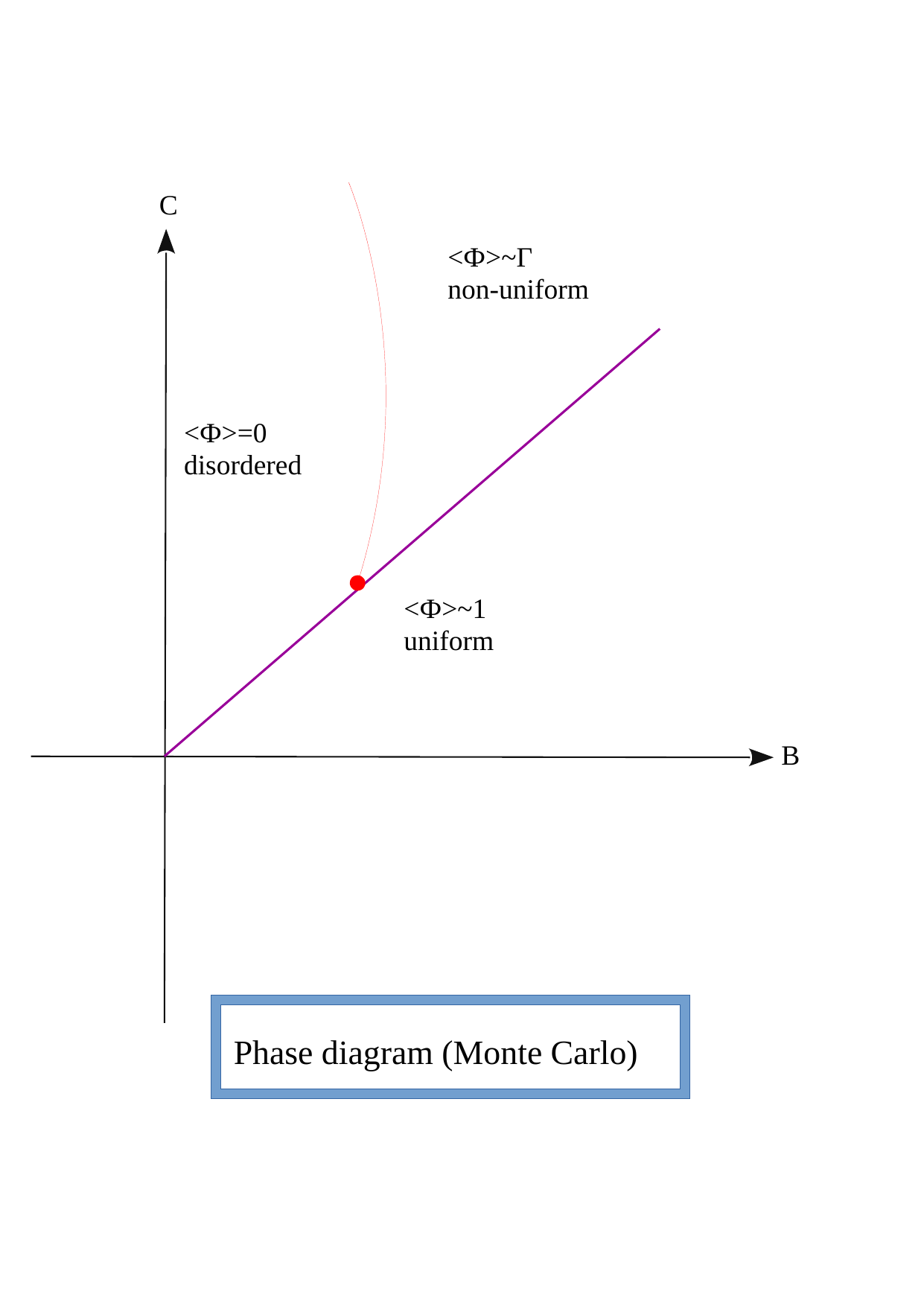}
\end{center}
\caption{The phase structure of the doubletrace cubic-quartic matrix model (\ref{BCD}) for values of $D$ consistent with noncommutative phi-four theory. $B$ is negative.}\label{sketch1}
%\caption{The continuum limit $\Lambda\longrightarrow \infty$ and the commutative limit $N\longrightarrow \infty$ of the eigenvalue distribution $\rho$ of the matrices $D_a$ for $\tilde{\alpha}=3$ across the critical boundary.}
\end{figure}

The phase structure is sketched on figure (\ref{sketch1}).
               
               We will employ repeatedly large $N$ factorization  of any multi-point function of $U(N)-$invariant objects ${\cal O}$,...,${\cal O}^{\prime}$ into product of one-point functions given by \cite{Higuchi:1994rv}
\begin{eqnarray}
  \langle {\cal O}...{\cal O}^{\prime}\rangle=\langle {\cal O}\rangle ....\langle{\cal O}^{\prime}\rangle+O(\frac{1}{N^2}).
\end{eqnarray}
               By expanding the doubletrace term and using large $N$ factorization  we obtain
               \begin{eqnarray}
                 Z&=&Z_{4}\sum_{n=0}\frac{1}{n!}(-\frac{g}{3})^n\langle (Tr_N M)^n(Tr_N M^3)^n\rangle_4\nonumber\\
                 &=&Z_{4}\sum_{n=0}\frac{1}{n!}(-\frac{g}{3})^n\big(\langle Tr_N M\rangle_4\big)^n\langle(Tr_N M^3)^n\rangle_4\nonumber\\
                 &=&Z_{4}\sum_{n=0}\frac{1}{n!}(-\frac{gNa_1}{3})^n\langle(Tr_N M^3)^n\rangle_4.\label{40}
               \end{eqnarray}
               The expectation value $\langle\rangle_4$ is computed with respect to the quartic potential $V_4=g_2M^2/2+g_4M^4/4$ and $Z_4$ is the corresponding partition function. Whereas $a_1$ is the first moment defined by $a_1=\langle Tr_NM/N\rangle$. In some sense this looks like a mean-field approximation.

               We obtain therefore from (\ref{40}) the cubic-quartic matrix model with coupling constants $g_3=ga_1$ and $g_4$, viz
               \begin{eqnarray}
                 &&Z=\int {\cal D}M~\exp(-N Tr_{N} V(M))\nonumber\\
                 &&V(M)=\frac{g_2}{2}M^2+\frac{g_3}{3}M^3+\frac{g_4}{4}M^4.
               \end{eqnarray}
             Next step is to expand the quartic term and use large $N$ factorization in a similar fashion together with Schwinger-Dyson identities. First, by expanding the quartic term and using large $N$ factorization  we obtain
               \begin{eqnarray}                 
                 Z&=&Z_{3}\sum_{n=0}\frac{1}{n!}(-\frac{Ng_4}{2g_3^2})^n\big(\frac{g_3^2}{2}  \langle Tr_N M^4\rangle_3\big)^n.\label{42}
               \end{eqnarray}
               Now, the expectation value $\langle\rangle_3$ is computed with respect to the cubic potential $V_3=g_2M^2/2+g_3M^3/3$ and $Z_3$ is the corresponding partition function. Second, we use  the $n=0$ and $n=1$ Schwinger-Dyson identities in the cubic potential given by
               \begin{eqnarray}
                 g_2\langle Tr_N M^2\rangle_3+g_3\langle Tr_N M^3\rangle_3=N.
               \end{eqnarray}
               \begin{eqnarray}
                 \langle Tr_N M\rangle_3=\frac{g_2}{2}\langle Tr_N M^3\rangle_3+\frac{g_3}{2}\langle Tr_NM^4\rangle_3.
               \end{eqnarray}
               By combining these two identities we can express $\langle Tr_N M^4\rangle_3$ in terms of $\langle Tr_N M\rangle_3$ and $\langle Tr_N M^2\rangle_3$. By substituting this expression into the partition function (\ref{42}) and using again large $N$ factorization (in reverse) we obtain
               
               \begin{eqnarray}                 
                 Z&=&Z_{3}\sum_{n=0}\frac{1}{n!}(-\frac{Ng_4}{2g_3^2})^n\langle\big(g_3  Tr_N M+\frac{1}{2}Tr_N M^2-\frac{Ng_2}{2}\big)^n\rangle_3\nonumber\\
                 &=&\int {\cal D}M~\exp(-N Tr_{N} V(M)).
               \end{eqnarray}
               The potential is given now by 
               \begin{eqnarray}
                 &&V(M)=\frac{g_4}{2g_3}M+(\frac{g_2}{2}+\frac{g_4}{4g_3^2})M^2+\frac{g_3}{3}M^3.
               \end{eqnarray}
               By shifting the field as $M\longrightarrow M+\delta$ and choosing $\delta$ in such a way that the linear term vanishes we obtain the cubic potential
  \begin{eqnarray}                 
                 V(M)=\frac{g_2^{\prime}}{2}M^2+\frac{g_3}{3}M^3~,~g_2^{\prime}=\pm\sqrt{(g_2+\frac{g_4}{g_3^2})^2-2g_4}.\nonumber\\
  \end{eqnarray}
  Thus, regardless of the sign of the initial $g_2$ we end up with a cubic potential with a positive (or negative) effective $g_2^{\prime}$. For simplicity, we take $g_2^{\prime}$ to be positive.

We know that the cubic potential admits the fixed points
  
  \begin{eqnarray}                 
                 \frac{g_3}{(g_{2}^{\prime})^{\frac{3}{2}}}\lvert_*=0.\label{result0}
  \end{eqnarray}
  \begin{eqnarray}                 
                 \frac{g_3}{(g_{2}^{\prime})^{\frac{3}{2}}}\lvert_*=\epsilon\equiv\frac{1}{432^{1/4}}.\label{result}
  \end{eqnarray}
  However, the original model depends on the two parameters $g_3$ and $g_4$. We can then solve (\ref{result}) for the critical value $g_{4*}$ as a function of $g_3$.
  The solution which goes through the two fixed points $(g_{3*},g_{4*})=(0,0)$ and  $(g_{3*},g_{4*})=(\epsilon,0)$ of the pure cubic potential is given explicitly by
  \begin{eqnarray}                 
                 g_{4*}=g_3^2\bigg[g_3^2-\sqrt{g_3^4-2g_2g_3^2+\big(\frac{g_3}{\epsilon}\big)^{4/3}}-g_2\bigg].\label{result1}
  \end{eqnarray}
This shows explicitly that (\ref{result}) is not just a single fixed point but a whole line of fixed points interpolating between the two points  $(g_{3*},g_{4*})=(0,0)$ and  $(g_{3*},g_{4*})=(\epsilon,0)$ in the plane $(g_3,g_4)$. Clearly, without the cubic-quartic interaction, there will be only one single point which is the Gaussian fixed point  $(g_{3*},g_{4*})=(0,0)$. And, for quartic interaction, there will also be one single fixed point in the positive quadrant in the plane $(g_3,g_4)$ given by this Gaussian fixed point  representing therefore the $3$rd order matrix transition. The extra fixed point should then be attached, as we will see below, with the Ising transition. Indeed, recall that the quartic term is our genuine phi-four interaction term whereas the cubic term represents, albeit approximately,  the effect of the kinetic term of a phi-four theory on a particular noncommutative space.  The inclusion of other multitrace terms such as $Tr M Tr M $, $Tr M^2 Tr M$, etc (with appropriate coefficients) should then be regarded as improving our approximation of the underlying kinetic term of a noncommutative phi-four theory and their treatment should go along the same lines.

  For $g_2<0$ the square root in the result (\ref{result1}) is well defined for all values of $g_3$. The behavior for $g_3^2\longrightarrow \infty$ and $g_3^2\longrightarrow 0$ is given by
   \begin{eqnarray}                 
                 g_{4*}=-\frac{1}{2}(\frac{g_3}{\epsilon})^{4/3}~,~g_3^2\longrightarrow \infty.
   \end{eqnarray}
   And
   \begin{eqnarray}                 
                 g_{4*}=g_3^2(-g_2-(\frac{g_3}{\epsilon})^{2/3})~,~g_3^2\longrightarrow 0.
   \end{eqnarray}
   Here $g_2=-1$ since $B$ is taken to be negative which is the region of interest for noncommutative scalar phi-four theories and their approximation with multitrace matrix models.

   In summary, the critical line (\ref{result}) interpolates smoothly between the two fixed points $(g_{3*},g_{4*})=(0,0)$ and  $(g_{3*},g_{4*})=(\epsilon,0)$ (going through a maximum in between) then it diverges to $g_{4*}\longrightarrow -\infty$ as $g_3\longrightarrow + \infty$ (which is an unphysical part of the critical line for noncommutative field theory).

   The critical line  (\ref{result}) (restricted to the positive quadrant) is naturally interpreted as the critical line  of fixed points associated with the phase transition between uniform and non-uniform-ordered phases. This line clearly interpolates between the fixed point $(0,0)$ (associated with the  transition between disordered and non-uniform-ordered phases) and the fixed point $(\epsilon,0)$ (associated with the transition between disordered and uniform-ordered phases).

   The transition between uniform and non-uniform-ordered phases is a noncommutative field theory phase transition, whereas the transition between disordered and non-uniform-ordered phases is a matrix model phase transition while the transition between disordered and uniform-ordered phases is a commutative geometry fixed point.

\begin{figure}[htbp]
\begin{center}  
  \includegraphics[width=10cm,angle=-0]{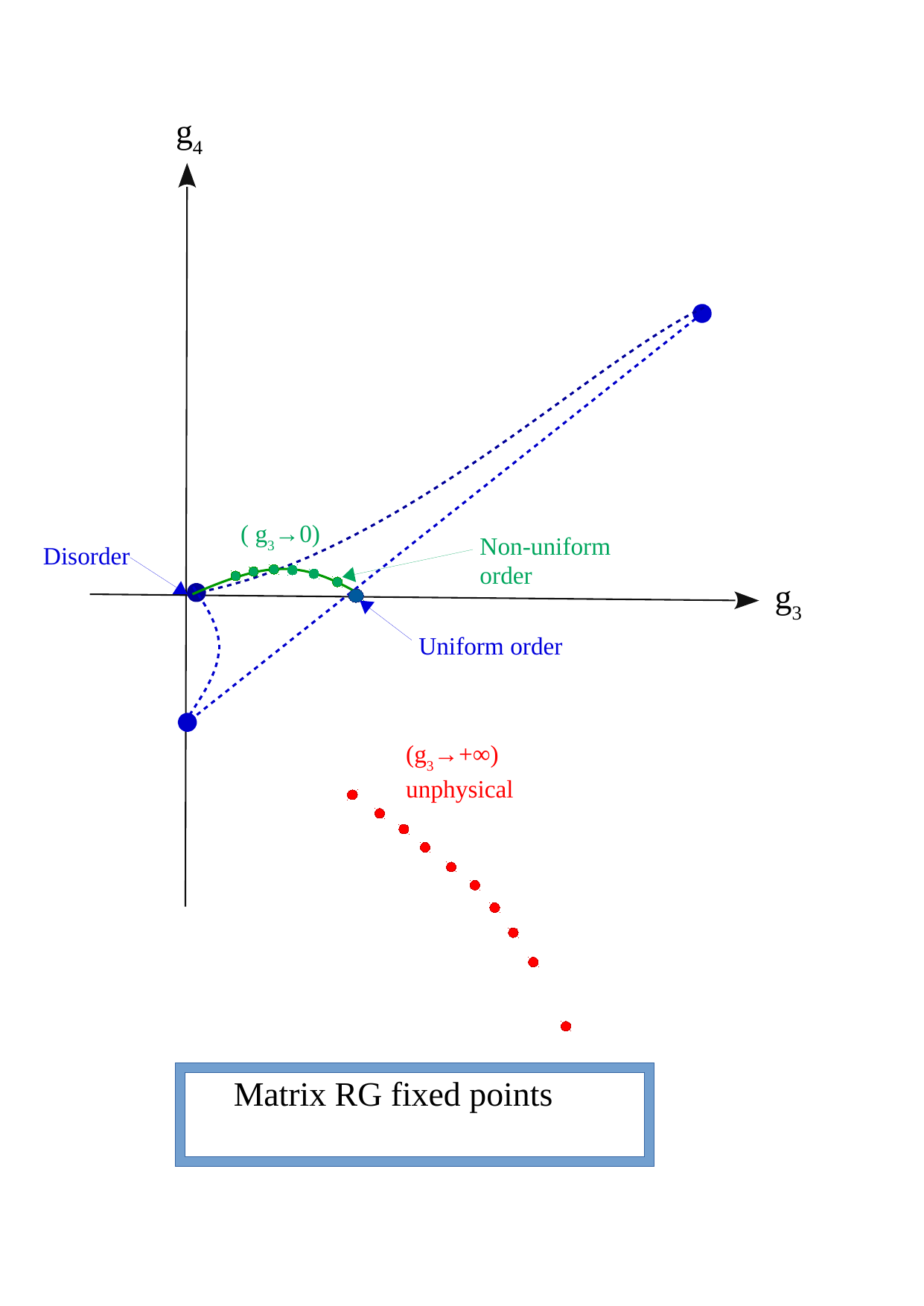}
\end{center}
\caption{The RG fixed points of the doubletrace cubic-quartic matrix model (\ref{BCD}) (green and red lines) compared with the RG fixed points of the singletrace cubic-quartic matrix model (blue line).}\label{sketch2}
%\caption{The continuum limit $\Lambda\longrightarrow \infty$ and the commutative limit $N\longrightarrow \infty$ of the eigenvalue distribution $\rho$ of the matrices $D_a$ for $\tilde{\alpha}=3$ across the critical boundary.}
\end{figure}

   In fact we believe that the  fixed point $(g_{3*},g_{4*})=(1/432^{0.25},0)$ corresponds to the  2nd order Ising phase transition (small values of $C$ and negative values of $B$), the fixed point  $(g_{3*},g_{4*})=(0.3066,0.0253)$ corresponds to the 2nd order phase transition between disordered and non-uniform-ordered phases (large values of $C$ and negative values of $B$), and the Gaussian fixed point $(g_{3*},g_{4*})=(0,0)$ corresponds to the 3rd order matrix phase transition. Naturally, negative values of $g_4$ are not of physical relevance to matrix models and multitrace matrix models of  noncommutative quantum field theory. The RG fixed points are sketched on figure (\ref{sketch2}).
   
\section{Conclusion}
It will be very interesting to generalize in a more rigorous way the Wilsonian matrix renormalization group equation outlined in this article to the general theory of multitrace matrix models  which are of great interest to  noncommutative quantum field theory and their phase structures as well as to models of emergent noncommutative geometry of quantum gravity. In particular, the phase structure of noncommutative phi-four remains a major question of paramount importance and the renormalization group approach together with the Monte Carlo method remain the two most important tools at our disposal in uncovering the non-perturbative physics induced by spacetime geometry.

\end{document}